\documentclass[final,times,5p]{elsarticle}
\usepackage{fancyhdr}

\usepackage{array}
\usepackage{gensymb}
\usepackage[section]{placeins}
\usepackage{caption}
\usepackage{float}

\makeatletter
\def\ps@pprintTitle{%
  \let\@oddhead\@empty
  \let\@evenhead\@empty
  \def\@oddfoot{\reset@font\hfil\footnotesize{\emph{\today}}}
  \let\@evenfoot\@oddfoot
}
\makeatother

\usepackage{siunitx}
\usepackage{inputenc}
\usepackage{cuted}
\usepackage{subfigure}
\usepackage[abs]{overpic}
\usepackage{booktabs}
\usepackage{textcomp}
\usepackage{multirow}
\usepackage{times}
\biboptions{numbers,sort&compress}
\newcolumntype{L}[1]{>{\raggedright\let\newline\\\arraybackslash\hspace{0pt}}m{#1}}
\newcolumntype{C}[1]{>{\centering\let\newline\\\arraybackslash\hspace{0pt}}m{#1}}
\newcolumntype{R}[1]{>{\raggedleft\let\newline\\\arraybackslash\hspace{0pt}}m{#1}}
\usepackage[bottom]{footmisc}

\usepackage{hyperref}
\usepackage{tabu}
\usepackage{enumitem}
\usepackage{scrextend}

\usepackage{booktabs, siunitx}
\usepackage[svgnames,table]{xcolor}
\usepackage[tableposition=below]{caption}
\usepackage{gensymb}
\usepackage{tikz}
\usetikzlibrary{arrows,calc,positioning}
\tikzstyle{B1}=[draw,text centered,minimum size=2.3 em,text width=4 cm,text height=0.1 cm]
\tikzstyle{B2}=[draw,text centered,minimum size=2.3 em,text width=1.5 cm,text height=0.1 cm]
\tikzstyle{B3}=[draw,text centered,minimum size=2.5 em,text width=2.5 cm,text height=0.1 cm]
\definecolor{mycolor}{RGB}{170,255,255}
\usepackage{graphicx}
\usepackage{amsmath}
\usepackage[nameinlink]{cleveref}
\crefname{figure}{Fig.}{Figs.}
\crefname{equation}{Eq.}{Eqs.}
\crefname{table}{Table}{Tables}
\crefname{section}{Section}{Sections}
\Crefname{figure}{Figure}{Figures}
\Crefname{equation}{Equation}{Equations}
\Crefname{table}{Table}{Tables}
\Crefname{section}{Section}{Sections}
\usepackage{amssymb}
\usepackage{setspace}
\usepackage{geometry}
\usepackage{longtable}
\usepackage{mathtools}
\usepackage{physics}
\usepackage{siunitx}
\usepackage{hyperref}
\hypersetup{
    colorlinks =    true,
    linkcolor =     [rgb]{0.0,0.0,1.0},
    anchorcolor =   [rgb]{0.0,0.0,1.0},
    citecolor =     [rgb]{0.0,0.0,1.0},
    filecolor =     [rgb]{0.0,0.0,1.0},
    urlcolor =      [rgb]{0.0,0.0,1.0},
    pdftitle=       {Title},
    pdfsubject=     {Title},
    pdfauthor=      {A. Uthor}
}
\numberwithin{equation}{section}

\newcommand{\drop}[1]{}
\begin{document}
\begin{frontmatter}
\title{Design of auxetic cellular structures for in-plane response through out-of-plane actuation of stimuli-responsive bridge films}

\author[inst1]{Anirudh Chandramouli\corref{cor2}}\ead{chandramoulianirudh@gmail.com}
\author[inst2]{Sri Datta Rapaka\corref{cor2}}\ead{datta5284@gmail.com}
\author[inst1]{Ratna Kumar Annabattula\corref{cor1}}
\ead{ratna@iitm.ac.in}
\cortext[cor2]{Equal contribution}
\cortext[cor1]{Corresponding author}
\affiliation[inst1]{organization={Department of Mechanical Engineering},
addressline={Indian Institute of Technology Madras}, 
city={Chennai},
postcode={600036}, 
state={Tamil Nadu},
country={India}}
\affiliation[inst2]{organization={Haas Formula One},
city={Banbury},
postcode={OX16 4PN},
country={UK}}
\begin{abstract}
    In this work, we propose novel designs of cellular structures exhibiting unconventional in-plane actuation responses to external stimuli. We strategically introduce stimuli-responsive bilayer bridge films within conventional honeycombs to achieve the desired actuation. The films are incorporated such that, in response to an external field (thermal, electric, chemical, etc.), the bridge film bends out-of-plane, activating the honeycomb in the plane. The conventional out-of-plane deformation of the bridge film can lead to interesting and unconventional actuation in the plane. An analytical model of this coupled unit cell behaviour is developed using curved beam theory, and the model is validated against finite element simulations. Several applications of such designs are presented. Unit cell architectures exhibiting both positive and negative macroscopic actuation are proposed, and the criterion for achieving such actuation is derived analytically. Furthermore, we demonstrate that by altering the topology, unidirectional and bidirectional negative actuation can be achieved. We also propose designs that result in the negative actuation of the structure with both monotonically increasing and monotonically decreasing stimuli. Finally, by combining two macroscopic structures with positive and negative actuation, we design efficient actuators/sensors that bend in the plane in response to a stimulus.
\end{abstract}

\end{frontmatter}

\section{Introduction}
Metamaterials are a class of materials that are engineered to exhibit unusual mechanical or physical properties not found in traditional materials \citep{barchiesi2019mechanical,surjadi2019mechanical,ren2018auxetic}. These materials can be designed to have specific properties that are desirable for various applications \citep{LONG2021113429,xin20224d,wang2021compression,cheng2022design}. One area of particular interest in the study of metamaterials is the development of materials with negative parameters of mechanical properties, which are not found in nature. Negative parameters of mechanical properties refer to values opposite in sign to what is typically observed in conventional materials. For example, negative thermal expansion (NTE) refers to a material contracting rather than expanding upon heating. While such behaviour is rare in nature, recent developments in lattice metamaterials have focused on optimizing mesoscale architecture to engineer materials which exhibit unusual macroscopic behaviour such as near zero expansion \citep{xie2018double,rhein2011bimetallic,yamamoto2014thin} and negative thermal expansion \citep{miller2009negative,n2007connected,lakes1996cellular,lakes2007cellular,grima2010composites,grima2015maximizing,takezawa2017design,wu2016isotropic,wang2016lightweight,peng2021novel,shen2022novel}. This counter-intuitive behaviour has found applications in various fields, including the development of temperature-stable components for aerospace and electronics applications \citep{chen2021stiffness,wei2018lightweight,wei2018three,lifson2017enabling,yang2016soft}. Similarly, metamaterials have been designed to obtain negative swelling behaviour \citep{ma20224d,zhang2018soft,curatolo2018effective,liu2016harnessing,chen2021hydrogel,wei2020design} and negative hygroscopic expansion \citep{ma2022design,bai2022moisture,lim2019composite,lim2023metamaterial,lim2021adjustable}, which have applications in the development of humidity-sensitive devices and materials.

Previous studies have achieved a negative (or reversed stimuli) response using one of two mechanisms. The first involves incorporating active components and clever kinematics to convert local axial expansions into macroscopic shrinkage. Examples of such designs include both two and three-dimensional structures formed using dual-material triangles \citep{wei2016planar,li2018hoberman,chen2022mechanical,huang2021auxetic} and dual-material tetrahedrons \citep{xu2018routes,xu2016structurally} respectively. The second mechanism involves introducing bi-material strips to replace certain members of the structure, resulting in bending and a negative response \citep{ha2015controllable,yu2019drastic,wu2016isotropic,wu2019mechanical}. While the former design offers high relative stiffness, it provides a limited range of negative expansion, whereas the latter offers a larger negative expansion but a lower stiffness.

To address the limitations of the two approaches mentioned above, we propose a novel design strategy that leverages out-of-plane deformation by strategically introducing stimuli-responsive bi-material bridge films into conventional lattice structures. This design allows for a greater macroscopic in-plane negative expansion range while maintaining high relative stiffness. 
Moreover, the actuation mechanism is out-of-plane, and only its projection into the plane governs the in-plane response, expanding the design space significantly. As such, our proposed design offers the capability to exhibit curious and unconventional behaviour, such as negative responses to both positive and negative changes in the stimulus, unidirectional or bidirectional negative expansion, etc. Moreover, we can achieve forward and reverse responses with a simple orientation change of the bridge film.

\section{Cellular structures with stimuli-responsive bridge films}
\begin{figure*}[h]
    \centering
    \includegraphics[width=0.9\textwidth,trim={0cm 0cm 0cm 0cm},clip]{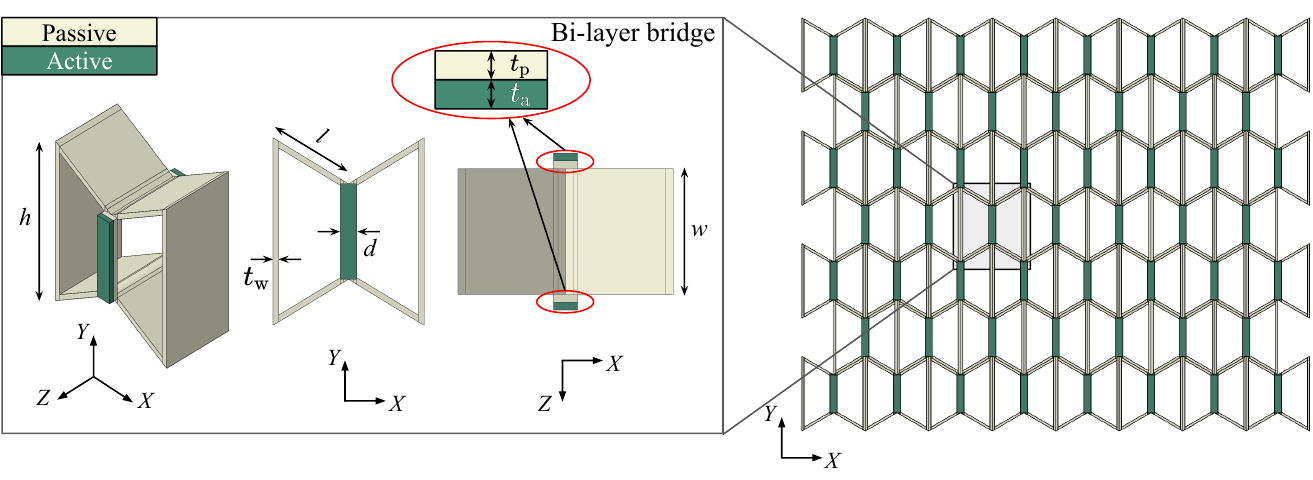}
    \caption{Schematic illustration of the unit cell of the proposed design with stimuli-responsive bilayer bridge films introduced between the vertices of a conventional auxetic honeycomb structure.}
    \label{fig:UnitCell}
 \end{figure*}

The proposed designs build on conventional cellular structures with well-developed in-plane mechanical behaviour. Stimuli-responsive bridge films are strategically introduced, connecting the vertices of the unit cells as shown in \cref{fig:UnitCell}. These bridge films are designed and oriented to deform in the out-of-plane direction. Accordingly, the bridge films comprise a stimuli-responsive active layer (which responds conventionally and expands in response to an increase in the external field) and a non-responsive passive layer stacked perpendicular to the plane of the unit cell. In response to an external stimulus, the bilayer bends out-of-plane due to a strain mismatch along its thickness, thereby inducing the unit cell in-plane actuation. Since only the ends of the films are connected to the vertices of the unit cell, the actuation of the unit cell depends on the projection of the film deformation onto the plane. Consequently, unconventional macroscopic in-plane responses to stimuli can be obtained through appropriate material properties and geometric considerations.


\subsection{Bridge film actuation} \label{sec:BFA}
The bilayer bridge film is assumed to respond to external stimuli such that the strain varies linearly with the external field. In other words, $\epsilon=\alpha\Delta T$, where $\Delta T$ is the applied field and $\alpha$ is the coefficient of expansion. The coefficient of expansion of the passive layer of the film and the unit cell is assumed to be negligibly small. The response of the bilayer film, in isolation, to an external field is well known \citep{timoshenko1925analysis}. With an increase in the applied field, the film both expands and bends out-of-plane with constant curvature along its length and is given by \citep{timoshenko1925analysis}
\begin{equation} \label{eqn:cdef}
    u=\frac{1}{R}=\frac{\alpha_\mathrm{a}-\alpha_\mathrm{p}}{\dfrac{t_\mathrm{p}+t_\mathrm{a}}{2}+\dfrac{(E_\mathrm{p} I_\mathrm{p} + E_\mathrm{a} I_\mathrm{a})}{t_\mathrm{p}+t_\mathrm{a}}\left(\dfrac{1}{E_\mathrm{p} t_\mathrm{p}}+\dfrac{1}{E_\mathrm{a} t_\mathrm{a}}\right)}\Delta T\triangleq c \Delta T,
\end{equation}
where $R$ is the radius of curvature of the film, $E_\mathrm{a}$ is the modulus of elasticity of the active layer, $t_\mathrm{a}$ is its thickness, $I_\mathrm{a}$ is its area moment of inertia, and $\alpha_\mathrm{a}$ is the coefficient of expansion. $E_\mathrm{p}$, $t_\mathrm{p}$, $I_\mathrm{p}$, and $\alpha_\mathrm{p}$ are the corresponding parameters of the passive layer. The projection of this deformation onto the plane, seen as a change in the in-plane length of the film, is given by
\begin{equation}
    \delta l=2 R \sin{\left(\frac{l_\mathrm{i}(1+\epsilon_\mathrm{p})}{2 R}\right)}-l_\mathrm{i}=\frac{\sin(l_\mathrm{i}[1+\epsilon_\mathrm{p}] c\Delta T/2)}{c \Delta T/2}-l_\mathrm{i},
\end{equation}
where $c$ is as defined in \cref{eqn:cdef}, $l_i$ is the initial length of the film and $\epsilon_p$ is the strain in the passive layer at the surface to be attached to the unit cell. This change can either be an increase or a decrease in length. Note that $\alpha_\mathrm{p} \to \alpha_\mathrm{a} \implies c \to 0$ and therefore $\delta l = l_\mathrm{i}\epsilon_\mathrm{p}$ resulting in a monotonic increase in the length with the applied field. However, if $\alpha_\mathrm{p} \neq \alpha_\mathrm{a}$, it is evident that $\delta l$ will eventually be negative for large $\Delta T$ resulting in an eventual in-plane shrinkage. However, to ensure an initial shrinkage,
\begin{equation}
    \frac{d\delta l}{d \Delta T}\bigg|_{\Delta T=0}<0\implies\alpha_\mathrm{p} \leq \left(\frac{5 E_\mathrm{a} E_\mathrm{p} - E_\mathrm{a}^2}{E_\mathrm{p} (19 E_\mathrm{a} + E_\mathrm{p})}\right) \alpha_\mathrm{a}.
\end{equation}
Note that for materials which do not satisfy the above relationship, the unit cell expands on heating (at least, initially). Therefore, by choosing the material parameters appropriately, both negative and positive actuation of the unit cell can be achieved in response to an increasing field. Since most materials expand with an increasing stimulus field, the more interesting case of negative response will be the primary focus of this work.



\subsection{Analytical model}
As discussed in \cref{sec:BFA}, the out-of-plane bending of the bridge film causes an in-plane change in its length. However, since the bridge film is attached to the vertices of the unit cell, the unit cell resists this change in length. Furthermore, since the film is assumed to be glued to the vertices, the unit cell also offers moment resistance to a change in the film angle at its ends. An analytical model is developed to study this coupled response. 

\begin{figure*}[htpb!]
\begin{center}
\subfigure[]{
    \includegraphics[width=0.55\textwidth]{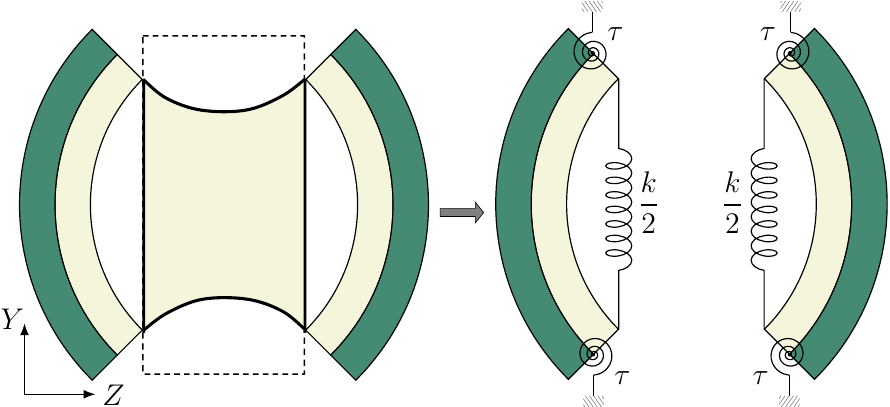}
    \label{fig:schemOfAna}
    }  \hfill
\subfigure[]{
    \includegraphics[width=0.39\textwidth]{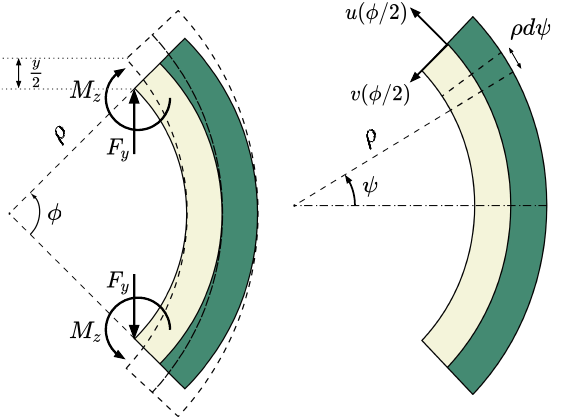}
    \label{fig:Ana}
    }
\end{center}
    \caption{(a) Schematic of the spring approximation of the unit cell. The translational response is captured by a linear spring whose stiffness is derived from \cite{gibson_ashby_1997}, and the torsional response is replaced by two torsional springs of stiffness, $\tau$. (b) Curved composite bilayer, initially at equilibrium under the combined action of a stimuli field differential and resisting moment, subject to spring forces and moments}
    \label{fig:AnalyticalModel}
 \end{figure*}

The bridge films at the inner and outer ends of the unit cells are assumed to be identical and mirror-symmetric about the X-Y plane as shown in \cref{fig:UnitCell}. Therefore, a half-cell model of the unit cell coupled with a single bridge film is developed. 
Furthermore, the in-plane mechanics of the conventional unit cell is modelled through a continuum approximation following the work of \citet{gibson_ashby_1997}. Additionally, assuming small strains (and geometry change), the in-plane mechanism is idealized as a translation spring (attached to the edges on the inner surface of the bilayer film) and two torsion springs (attached to both ends of the film and, in turn, coupled to the ground). The schematic of this simplification is shown in \cref{fig:schemOfAna}. The stiffness of the translation spring, $k$, for a hexagonal honeycomb is given as \citep{gibson_ashby_1997}
\begin{equation}
    k=\frac{E_\mathrm{h} w t_w^3}{l^3 \cos^2{\theta}},
\end{equation}
where $l$ is the inclined wall length as shown in \cref{fig:UnitCell}, $w$ is the out-of-plane width, $t_w$ is the wall thickness, $\theta$ is the inclination angle, and $E_\mathrm{h}$ is the Young's modulus of the unit cell material. Similarly, $\tau$ is the stiffness of the equivalent torsional spring. An approximation of $\tau$ can be found by assuming that the deformations caused due to moment are in the form of suitable polynomials. The coefficients are subsequently found using strain energy minimization and, in turn, used to derive $\tau$. The details of the derivation and the analytical expression for $\tau$ are presented in the supplementary file\footnote{The MATHEMATICA code used for the derivation of an expression for $\tau$ can be downloaded from \href{https://github.com/AnirudhChandramouli/HexagonalUnitCell\_Stiffness.git}{https://github.com/AnirudhChandramouli/HexagonalUnitCell\_Stiffness.git}}.

Using the above simplifications, the response of the coupled system is found as follows. Firstly, the bending moment arising due to the strain mismatch in the bilayer and the resisting moment offered by the torsional springs are solved for equilibrium. It is assumed that the total deformation of the system can be seen as a superposition of this pure bending deformation and the deformations arising from the force in the translational spring. 

Firstly, we derive the equilibrium configuration of the bilayer and torsional springs. Since the system is such that the bending moment will be constant through the length, the curvature will also be constant. Let the corresponding radius of curvature at the equilibrium be $\rho$ and the subtended angle be $\phi$. Then, we have
\begin{equation} \label{rho1}
    \rho = \frac{E_\mathrm{c} I_\mathrm{c}}{M}=\frac{E_\mathrm{c} I_\mathrm{c}}{\tau \phi + M_\mathrm{th}}        
\end{equation}
where $E_\mathrm{c}$ and $I_\mathrm{c}$ are the effective Young's modulus and area moment of inertia of the composite bilayer, respectively and $M_\mathrm{th} = E_\mathrm{c} I_\mathrm{c}/R$ is the equivalent thermal moment. Since the composite bar is in pure bending, the length of the neutral axis remains constant. Therefore, we have $\rho \phi = R \theta$. Upon substitution of $\rho={R\theta}/{\phi}$ in \cref{rho1}, we have
\begin{align} 
    &\frac{R\theta}{\phi} =\frac{E_\mathrm{c} I_\mathrm{c}}{\tau \phi + M_{th}} \\
    &\tau R\theta + \frac{E_\mathrm{c} I_\mathrm{c}\theta}{\phi} =E_\mathrm{c} I_\mathrm{c}\\
    &\tau R\theta \phi + E_\mathrm{c} I_\mathrm{c}\theta = E_\mathrm{c} I_\mathrm{c}\phi \\
    \implies &\phi = \frac{E_\mathrm{c} I_\mathrm{c}\theta}{E_\mathrm{c} I_\mathrm{c} - \tau R\theta}\\
    &\rho = \frac{E_\mathrm{c} I_\mathrm{c} R - \tau R^2 \theta}{E_\mathrm{c} I_\mathrm{c}} \label{eqn:eqConf}
\end{align}
This equilibrium configuration, characterized by $\rho$ and $\phi$, is assumed to be the initial configuration of a curved beam which is then subjected to a splitting force ($F_y$) and moment ($M_z$) as shown in \cref{fig:Ana}. Analyzing the beam in polar coordinates and considering an infinitesimal segment $d\psi$ at an angle of $\psi$ from the centre, we have \cite{oden1982mechanics}
\begin{align}
    \frac{1}{\rho} \frac{du}{d\psi}-\frac{v}{\rho}&=\frac{-F_y \rho\cos\phi-M_z}{\rho A_\mathrm{c}E_\mathrm{c}},  \label{eqn:u}\\
    \frac{1}{\rho^2} \frac{d^2v}{d\psi^2}+\frac{v}{\rho^2}&=\frac{F_y\rho(\cos{\psi}-\cos\phi)-M_z}{E_\mathrm{c}I_\mathrm{c}} \label{eqn:v},
\end{align}
where $A_\mathrm{c}=(A_\mathrm{p}E_\mathrm{p}+A_\mathrm{a}E_\mathrm{a})/E_\mathrm{c}$ is the effective area of the composite bilayer, $u$ and $v$ are the tangential and radial displacements respectively. The above equations are solved with the boundary conditions $v|_{\psi=0}=0$, $\frac{1}{\rho} \frac{dv}{d\psi}\bigg|_{\psi=0}=0$, and $u|_{\psi=0}=0$ that represent the symmetric deformation of the curved composite beam, resulting in
\begin{align}
    u(\psi)&=\frac{1}{4 \rho}\bigg(\frac{\rho^3(2  \rho F_y \sin{2\psi}+(4 M_z + 2  \rho F_y)\sin{\psi}-4 M_z \psi}{E_\mathrm{c} I_\mathrm{c}} \nonumber \\
    &\quad-\frac{6 \rho F_y \psi\cos{\psi}}{E_\mathrm{c} I_\mathrm{c}} - \frac{4\psi(M_z+\rho^2 F_y \cos{\psi})}{A_\mathrm{c} E_\mathrm{c}}\bigg),\\
    v(\psi)&=\frac{\rho^2}{4 E_\mathrm{c} I_\mathrm{c}}\bigg(4 \rho F_y \cos^2{\psi}+4(M_z-\rho F_y)\cos{\psi}-4 M_z \nonumber \\
    &\quad+2 \rho F_y \psi \sin{\psi}\bigg),\\
    d\phi(\psi)&=\frac{\rho}{E_\mathrm{c} I_\mathrm{c}}(M_z \psi + \rho F_y \psi \cos{\psi} - \rho F_y \sin{\psi}).
\end{align}
Therefore, the vertical deformation ($y$) of the curved beam from the equilibrium configuration is given by
\begin{equation} \label{eqn:1}
    y=2 u(\phi/2) \cos{(\phi/2)} - 2 v(\phi/2) \sin{(\phi/2)}.
\end{equation}
Note that the external force ($F_y$) and moment ($M_z$) can only arise from the reaction forces from the translational and torsional springs. Therefore, the translational deformation ($y$) and angular deformation ($d\phi$) arising from these spring forces and moments must be compatible with the spring equations. Thus, we have
\begin{equation} \label{eqn:2}
    F_y=-k \delta^* \quad \text{ where } \delta^*=l_\mathrm{i}-2 \rho \sin(\phi/2)-y
\end{equation}
\begin{equation} \label{eqn:3}
    M_z=-\tau d\phi
\end{equation}
Note that $M_z$ is the moment on the bridge film over the equilibrium configuration defined by \cref{eqn:eqConf} and not the total external moment on the bridge film. The above system of equations (\cref{eqn:1,eqn:2,eqn:3}) are solved for displacement and forces in the spring. Accordingly, the vertical deformation of the unit cell as a function of temperature is given by  
\begin{align}
    \delta^*&=l_\mathrm{i}-2 \rho \sin(\phi/2)\nonumber\\
    & -\frac{k \rho^3 (l_\mathrm{i}-2 \rho \sin(\phi/2)) (\rho \tau \beta + 2 E_\mathrm{c}I_\mathrm{c} (2 \phi - \alpha))}{k \rho^4 \tau \beta + 2 E_\mathrm{c}I_\mathrm{c}(\rho \phi(2 k \rho^2+\tau)+ k \rho^3(\alpha+2 E_\mathrm{c}I_\mathrm{c}))},
\end{align}
where $\alpha=\phi \cos{\phi}-3 \sin{\phi}$ and $\beta=(-4+\phi^2+4 \cos{\phi}+\phi\sin{\phi})$. Here, $\delta^*$ is the total vertical deformation of the combined system from the initial configuration.
Further, the vertical force applied by the bilayer on the unit cell is given by
\begin{equation}
   F_y=-\frac{2 E_\mathrm{c}I_\mathrm{c} k (l-2 \rho \sin(\phi/2))(\rho \phi \tau + 2E_\mathrm{c}I_\mathrm{c})}{k \rho^4 \tau\beta+ 2E_\mathrm{c}I_\mathrm{c}(\rho \phi(2 k \rho^2+\tau)+k\rho^3\alpha+2E_\mathrm{c}I_\mathrm{c})}.
\end{equation}
Similarly, the corresponding moment applied is given by
\begin{equation}
   M_z=-\frac{2 k \rho^2 \tau (\phi\cos(\phi/2)-2\sin(\phi/2))(l-2 E_\mathrm{c}I_\mathrm{c} \rho \sin(\phi/2))}{k \rho^4 \tau\beta+ 2E_\mathrm{c}I_\mathrm{c}(\rho \phi(2 k \rho^2+\tau)+k\rho^3\alpha+2E_\mathrm{c}I_\mathrm{c})}.
\end{equation}
\subsection{Validation of the analytical model}
The developed analytical model is validated against finite element simulations of the unit cell. A single unit cell of the metamaterial (with the bridge films) is modelled in isolation, and its response is simulated. Both the unit cell and the bilayer are modelled using shell elements with reduced integration in Abaqus~\cite{simulia2012abaqus}. For the purpose of demonstration, the unit cell and the passive layer are assumed to be made of the same material. The geometric and material properties of the proposed unit cell are as defined in \cref{tab:MaterialProps}. The external stimulus is applied in the form of a temperature field to the model. Boundary conditions are applied, constraining the left surface of the unit cell in the $X$ direction and a point on the surface in the $Y$ and $Z$ directions to prevent rigid body motions. The response of the system is compared with the analytical predictions. 

\begin{table*}[h]
\centering
\caption{Geometric and material properties used in the model}
\label{tab:MaterialProps}
\begin{tabular}{p{8cm}l}
\hline
Young's modulus of active material ($E_\mathrm{a}$) & \SI{3} {\GPa} \\ \hline
Young's modulus of passive material ($E_\mathrm{p}$) & \SI{3} {\GPa} \\ \hline
Poisson's ratio ($\nu_\text{a},\nu_\text{p}$) & $0$ \\ \hline
Coefficient of expansion of active layer ($\alpha_\mathrm{a}$) & \SI{2e-3} {\per \degreeCelsius} \\ \hline
Coefficient of expansion of passive layer ($\alpha_\mathrm{p}$) & \SI{1e-10} {\per \degreeCelsius}\\ \hline
Inclined wall length ($l$) & \SI{4} {\mm} \\ \hline
Vertical wall length ($h$) & \SI{8} {\mm} \\ \hline
Hexagonal inclination angle ($\theta$) & \SI{-30}{\degree} \\ \hline
Cell wall thickness ($t_\text{w}$) & \SI{0.13} {\mm} \\ \hline
Width of the bilayer bridge film ($d$) & \SI{0.8} {\mm} \\ \hline
Out-of-plane width of the unit cell ($w$) & \SI{4} {\mm} \\ \hline
Thickness of the bilayer bridge film ($t_\text{a}, t_\text{p}$) & \SI{0.25} {\mm} \\ \hline
\end{tabular}
\end{table*}

\begin{figure*}[htpb]
    \centering
    \includegraphics[width=0.8\textwidth,trim={0cm 0cm 0cm 0cm},clip]{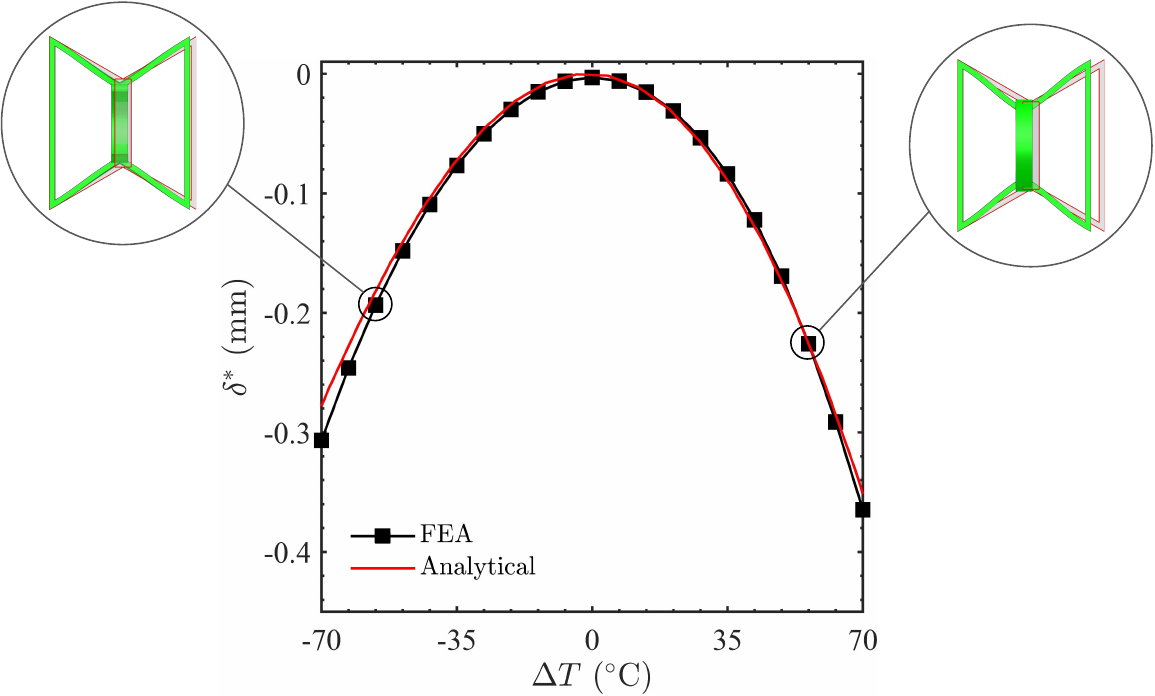}
    \caption{Comparison of the FEA results with analytical predictions with respect to the vertical deformation of the proposed unit cell under a temperature differential.}
    \label{fig:TempPlots}
 \end{figure*}

The vertical in-plane shrinkage ($\delta^*$) of the bilayer from the FEA simulations is plotted against the applied temperature field in \cref{fig:TempPlots} along with the analytical predictions. It is evident that the analytical model accurately captures the response of the unit cell over a large domain. This shrinkage of the bilayer actuates the unit cell by pulling the vertices of the hexagon together. Note that while the vertices are drawn together, the length of the side walls remains unaffected. Moreover, it is evident that the response is not symmetric for positive and negative fields. This is due to two reasons. Firstly, although the curvature of the film depends only on the magnitude of the applied field, the subtended angle (and consequently the change in length) depends on the sign of the applied field. Secondly, and more significantly, the thickness affects the response since the film is attached to the unit cell on its inner surface. During bending, the strain on the inner and outer surfaces of the film are not identical. This effect, however, diminishes as the ratio of length to thickness grows large.
The horizontal response of the unit cell is due to the effective Poisson effect of the metamaterial. The Poisson's ratio of a hexagonal unit cell is given by \citep{gibson_ashby_1997}
\begin{equation}\nu_{YX}=\frac{(h/l+\sin{\theta})\sin{\theta}}{\cos^2{\theta}}.
\end{equation}
For the auxetic structure chosen, the negative Poisson's ratio implies that the vertical shrinkage results in a shrinkage along the horizontal direction. The deformed configuration of the unit cell at $\pm \SI{50}{\degreeCelsius}$ is overlaid in \cref{fig:TempPlots}. It is interesting to note that both increasing and decreasing fields result in a shrinkage of the unit cell.


\section{Results and discussions}
In the previous section, a unit cell model was proposed that shrinks in response to an increasing field. Cellular structures with varied macroscopic behaviours can be engineered using these cells as building blocks. In this section, several interesting applications of such designs are discussed.

\subsection{Bi-directional negative actuation}
\begin{figure*}[htpb!]
\begin{center}
\subfigure[]{
    \includegraphics[width=0.54\textwidth]{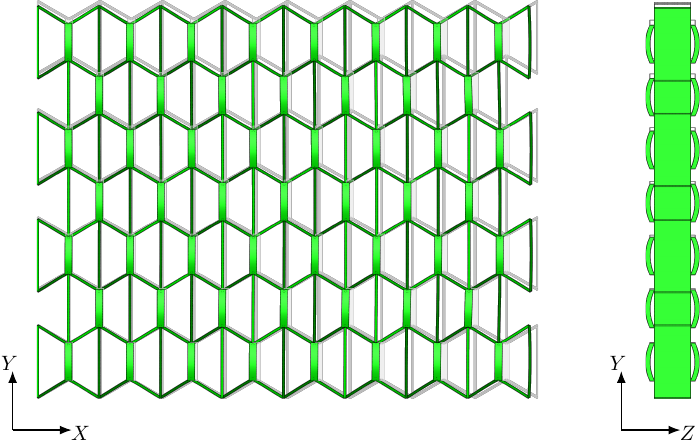}
    \label{fig:FV}
    }   \hfill
\subfigure[]{
    \includegraphics[width=0.34\textwidth]{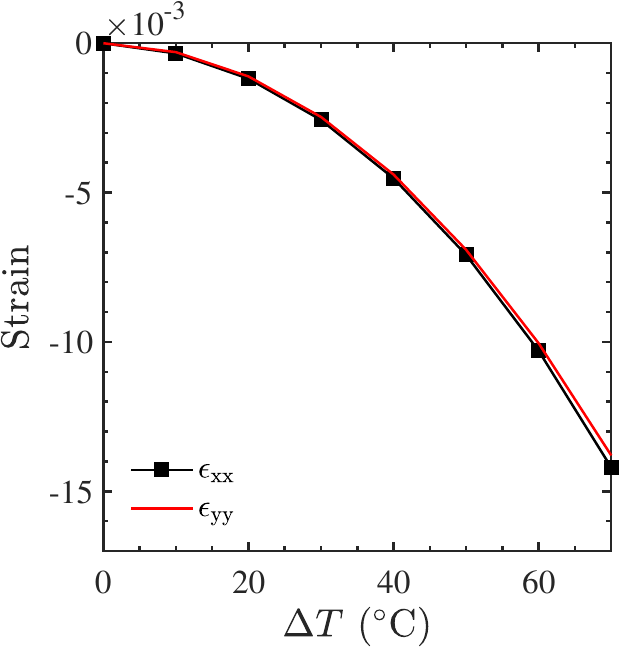}
    \label{fig:SV}
    }
\end{center}
    \caption{(a) Finite element simulations of the bi-directional negative expansion of the cellular structure under a temperature differential showing the deformed (green) and undeformed (grey) configuration at $\SI{70}{\degreeCelsius}$ and (b) the corresponding macroscopic strain variation in the bulk of the lattice structure.}
    \label{fig:BiDir}
 \end{figure*}

As noted previously, the bridge film pulls the mid-vertices of the hexagonal unit cell closer together. The length of the side walls remains unchanged, but the horizontal distance between them shrinks as a consequence of the Poisson effect. A row of such cells will each deform individually, and their total horizontal deformation will be a sum of the individual deformations. In stacking these cells vertically, each row can be arranged such that the side walls of any row are attached to the bridge films above and below it, as shown in \cref{fig:FV}. Consequently, each column is made of alternating side walls and bridge films. The total vertical deformation is, therefore, approximately equal to $n \delta^*/2$, where $n$ is the number of columns. However, in arranging the rows as such, the side walls attached above and below the bridge films add additional moment resistance (in addition to $\tau$). This stiffens the unit cell and reduces the vertical shrinkage (and, in turn, the horizontal shrinkage). 

\Cref{fig:FV} presents the undeformed and deformed configuration of a $7\times8$ cellular structure modelled as described above. The deformed configuration is plotted for a temperature field of $\SI{70}{\degreeCelsius}$. \Cref{fig:SV} presents the macroscopic strain variation in the bulk of the structure as a function of temperature. It is evident that the macroscopic response of the cellular structure is a bi-directional contraction with an increasing external field. Further, the strain along the horizontal and vertical at $\SI{70}{\degreeCelsius}$ are found to be 0.0144 and 0.0146, respectively. 

\subsection{Uni-directional negative actuation}
\begin{figure*}[htpb!]
\begin{center}
\subfigure[]{
    \includegraphics[width=0.553\textwidth]{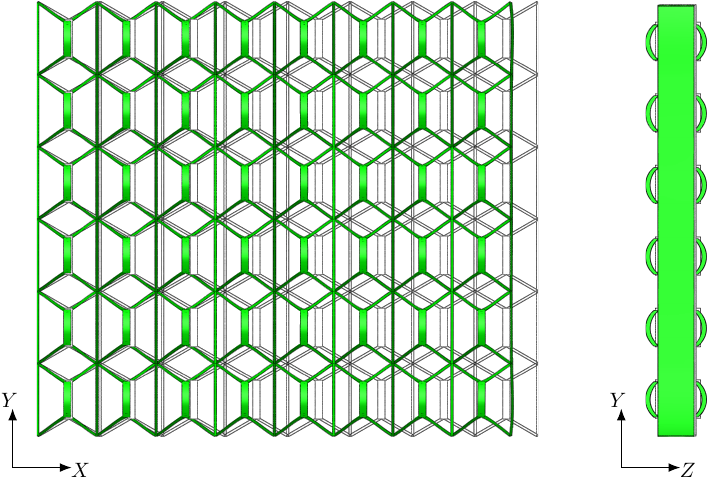}
    \label{fig:UFV}
    }    \hfill
\subfigure[]{
    \includegraphics[width=0.37\textwidth]{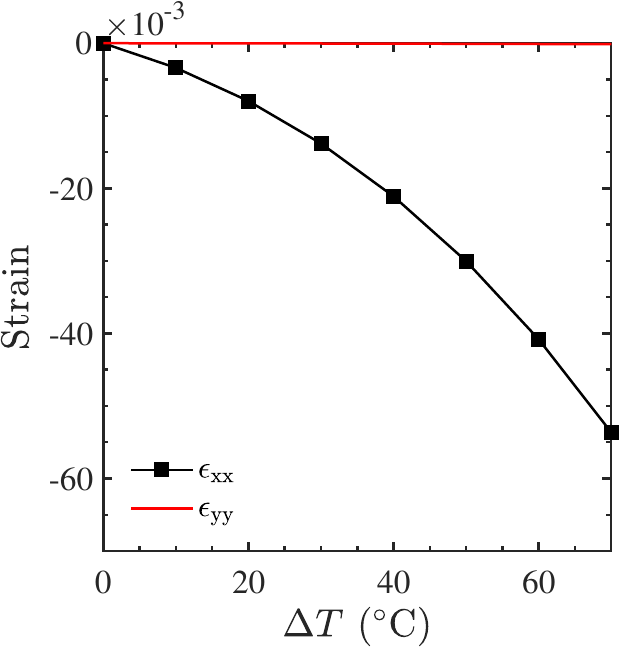}
    \label{fig:USV}
    }
\end{center}
    \caption{(a) Finite element simulations of the uni-directional negative expansion of the cellular structure under a temperature differential showing the deformed (green) and undeformed (grey) configuration at $\SI{70}{\degreeCelsius}$ and (b) the corresponding macroscopic strain variation in the bulk of the lattice structure.}
    \label{fig:UniDir}
 \end{figure*}

Since the side walls do not deform vertically, the rows of unit cells can be stacked such that each side wall connects to the corresponding wall of the rows above and below, as shown in \cref{fig:UFV}. This implies that the vertices remain unconnected and do transmit their deformations vertically. Consequently, the height of the cellular structure remains unchanged. However, due to Poisson's effect, the structure shrinks horizontally. Therefore, the structure actuates unidirectionally in response to an external stimulus. The deformed and undeformed configuration of a cellular structure with the above design is simulated in Abaqus and plotted in \cref{fig:UFV}. \Cref{fig:USV} presents the corresponding macroscopic strain variation in the bulk of the structure as a function of temperature. Since the vertices are unconnected, there is no increase in the moment stiffness of the unit cells. This consequently leads to a larger actuation, and the horizontal strain is consequently greater for this case, while the macroscopic vertical strain is $\approx 0$. At $\SI{70}{\degreeCelsius}$, the horizontal strain is found to be 0.054.

\subsection{In-plane bridge film actuation}
\begin{figure*}[htpb!]
\begin{center}  
\subfigure[]{
    \includegraphics[width=0.55\textwidth]{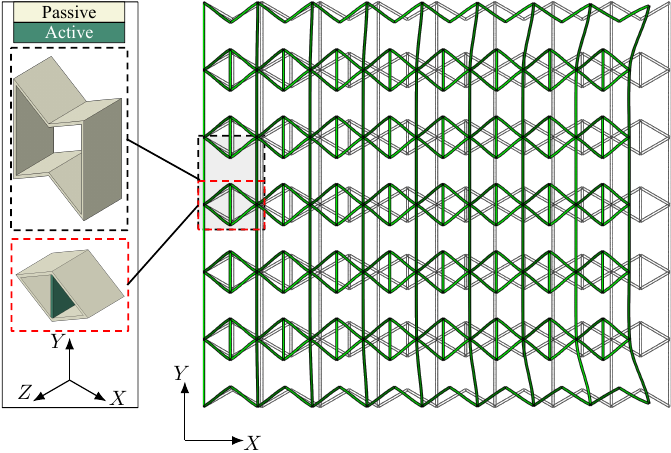}
    \label{fig:IPFV}
    }    \hfill
\subfigure[]{
    \includegraphics[width=0.37\textwidth]{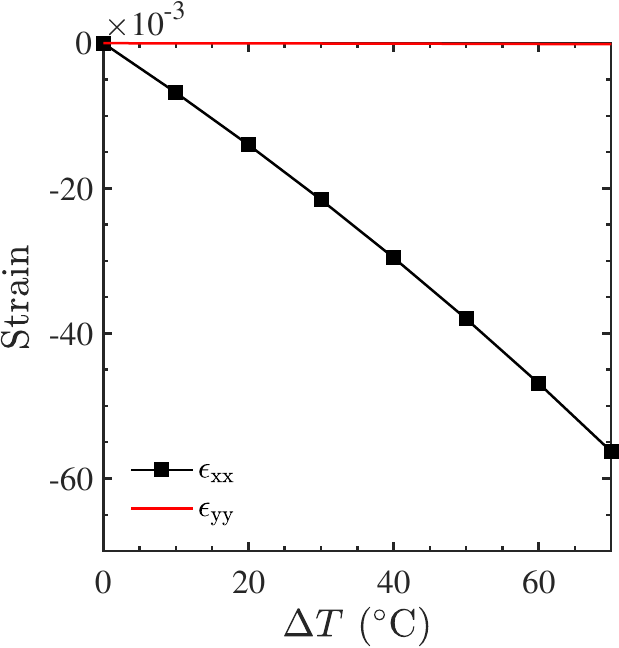}
    \label{fig:IPSV}
    }
\end{center}
    \caption{(a) Finite element simulation of the uni-directional negative expansion of the cellular structure under a temperature differential using in-plane bridge film actuation showing the deformed (green) and undeformed (grey) configuration at $\SI{70}{\degreeCelsius}$ and (b) the corresponding macroscopic strain variation in the bulk of the lattice structure.}
    \label{fig:InPlane}
 \end{figure*}

In the above design, vertical deformations were not transmitted since the vertices remained unconnected. In other words, the columns containing the stimuli-responsive bridge films were intermittent. Taking inspiration from the above mechanism of isolating the vertical deformations, these columns can be inverted - removing the bridge films within the unit cell and instead using bridge films connecting the vertices of one unit cell to that of the cell below and above it. Due to this inversion, an expanding film is required to achieve the same macroscopic response. Therefore, the bridge film could be introduced in-plane as a single active layer as shown in \cref{fig:IPFV}. This design can achieve a similar unidirectional negative actuation in response to an increasing stimulus field. The actuation response of this design is modelled and simulated in Abaqus. The deformed and undeformed configurations are plotted in \cref{fig:IPFV}, and the macroscopic strain variation as a function of temperature is plotted in \cref{fig:IPSV}. It is evident that the local vertical expansions result in a macroscopic shrinkage along the horizontal, with the macroscopic height of the structure remaining unchanged. Moreover, the strain is slightly larger and is found to be 0.056 at $\SI{70}{\degreeCelsius}$. However, it should be noted that the relative volume of the active material is significantly higher in this design. Further, it is noted from \cref{fig:USV} and \cref{fig:IPSV} that the out-of-plane design results in a nonlinear response in contrast to the in-plane design.

\subsection{Application as a bending actuator}
\begin{figure*}[htpb!]
\begin{center}
\subfigure[]{
    \includegraphics[width=0.4\textwidth]{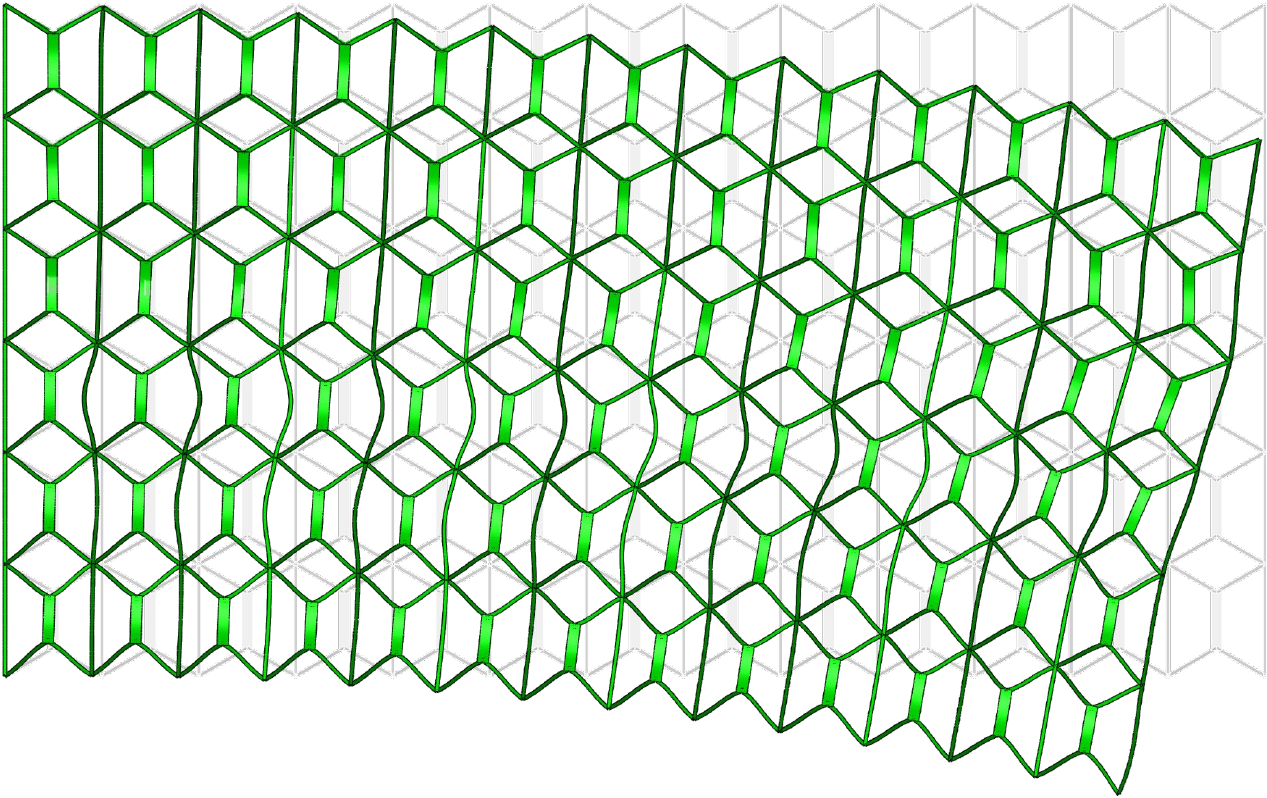}
    \label{fig:CaseA}
    }    \hspace{0.1\textwidth}
\subfigure[]{
    \includegraphics[width=0.4\textwidth]{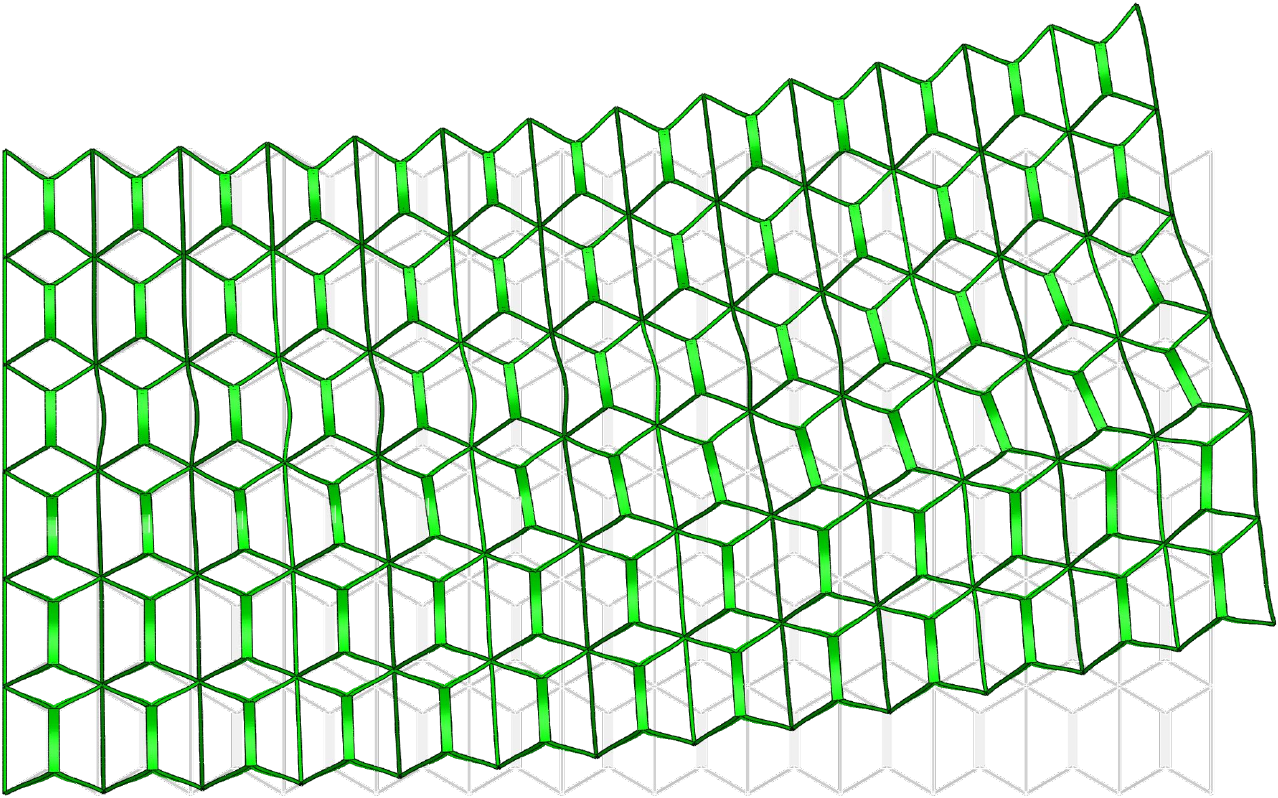}
    \label{fig:CaseB}
    }
\end{center}
    \caption{Bending of the bilayer lattice with its upper rows consisting of reversed response unit cells and its lower rows comprising regular response unit cells under (a) a negative temperature differential and (b) a positive temperature differential. The undeformed and deformed configurations are shown in grey and green, respectively.}
    \label{fig:Bilayer}
 \end{figure*}

The analytical model can be used to tune individual unit cells to achieve a varied set of responses. For instance, it can be shown that swapping the active and passive layers of the bridge film in the unit cell defined in \cref{tab:MaterialProps} leads to an initial expansion of the vertices. This consequently results in a macroscopically positive thermal expansion. Such unit cells can be combined with macroscopic negative expansion unit cells to engineer a differential in the macroscopic strains of the different layers. This consequently results in the bending of the lattice structure. Moreover, since the curvature is directly proportional to the difference in the coefficients of thermal expansions of the two layers, more efficient bending actuation can be generated since the coefficient can be tuned to be highly negative as opposed to regular materials, wherein the expansion coefficient is typically positive. The response of such an arrangement to an applied external stimulus is shown in \cref{fig:Bilayer}. The structure bends with both decreasing (\cref{fig:CaseA}) and increasing (\cref{fig:CaseB}) stimulus fields, but in opposite directions.

\section{Conclusion}

In this work, we present a new design strategy that enables the creation of lattice structures with unconventional in-plane responses. The proposed approach involves incorporating stimuli-responsive bilayer bridge films into conventional lattice structures. These films are oriented in a way that directs the deformations out-of-plane, allowing for the introduction of actuators with a large design space while having minimal impact on the in-plane characteristics of the base lattice structure.
An analytical model is developed using curved beam theory to better understand the response of unit cells comprising a conventional honeycomb and a stimuli response bridge film. This model considers the coupled response of the two systems, which can lead to complex and counter-intuitive behaviour. The analytical model is validated through finite element simulations, which confirm the accuracy of the model.
Potential applications of such designs are demonstrated through various finite element simulations. For instance, it is shown that the proposed design strategy can lead to highly sensitive bending actuators, which can be useful in applications such as sensors and soft actuators in robotics. Moreover, directional stimuli-responsive actuators are also demonstrated, which can be useful in applications such as soft robots or adaptive structures. Such lattice structures may find utility in various industries, such as aerospace and electronic applications. In summary, the proposed design strategy is a promising approach to engineer lattice structures with unconventional in-plane responses. The analytical model and finite element simulations demonstrate the feasibility of the proposed approach and its potential applications.

\bibliography{references}

\end{document}


\begin{frontmatter}
\title{Supplementary Material:\\Design of auxetic cellular structures for in-plane response through out-of-plane actuation of stimuli-responsive bridge films}
\author[inst1]{Anirudh Chandramouli\corref{cor2}}
\author[inst2]{Sri Datta Rapaka\corref{cor2}}
\author[inst1]{Ratna Kumar Annabattula}
\cortext[cor2]{Equal contribution}
\affiliation[inst1]{organization={Department of Mechanical Engineering},
addressline={Indian Institute of Technology Madras}, 
city={Chennai},
postcode={600036}, 
state={Tamil Nadu},
country={India}}
\affiliation[inst2]{organization={Haas Formula One},
city={Banbury},
postcode={OX16 4PN},
country={UK}}
\end{frontmatter}
\hrule
\section{Derivation of equivalent stiffness of torsion spring for a hexagonal honeycomb unit cell}
The analytical study of the coupled response of a unit cell and an attached stimuli-responsive bridge film to an external stimuli field in this work involved approximating the unit cell response to a linear spring and two torsion springs. While the derivation of equivalent linear response to forces is available in the literature \citep{gibson_ashby_1997}, the study of its response to moments at its vertex is not available; therefore, the equivalent torsional spring stiffness is not readily known. Therefore, in this section, the derivation of this stiffness is discussed\footnote{The MATHEMATICA code used for the derivation of an expression for $\tau$ can be downloaded from \href{https://github.com/AnirudhChandramouli/HexagonalUnitCell_Stiffness.git}{https://github.com/AnirudhChandramouli/HexagonalUnitCell\_Stiffness.git}}. 
\begin{figure}[htpb]
\begin{center}
\subfigure[]{
    \includegraphics[width=0.38\textwidth]{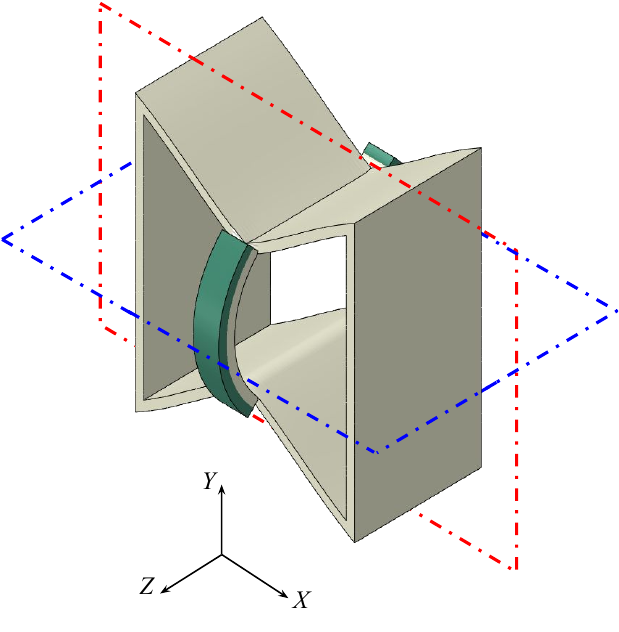}
    \label{fig:Schem1}
    } \hfill
\subfigure[]{
    \includegraphics[width=0.53\textwidth]{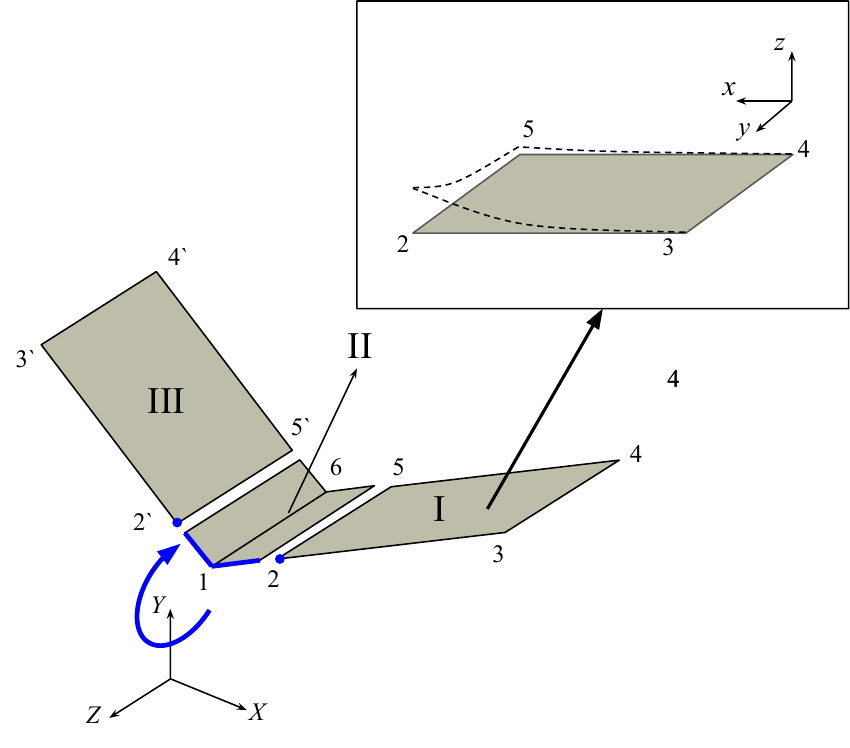}
    \label{fig:Schem2}
    } 
\end{center}
    \caption{(a) Hexagonal unit cell subject to moments at the vertices due to the bending of the attached stimuli responsive bridge film. The red and blue planes indicate planes of symmetry. (b) Simplified model of the response of a quarter section of the bridge film to an external moment about the $X$-axis at the vertex.}
    \label{fig:Schematic}
\end{figure}

Consider the unit cell subject to a moment at its vertices, arising from the bending of the bridge film, as shown in \cref{fig:Schem1}. Note that the resulting moment acts only at the vicinity of the vertex and thus cannot be analyzed as a simple bending problem of the entire unit cell section. Consequently, to derive its stiffness with respect to this moment, the system is idealized as presented in \cref{fig:Schem2}. The vertical side walls show negligible deformations, and the inclined walls are split into three sections. Section II is the section that is directly attached to the bridge film, and sections I and III are mirror symmetric. The following assumptions are made in the derivations which follow
\begin{itemize}
    \item The total stiffness of the unit cell is seen as the stiffness of the three sections acting in parallel (additive).
    \item The wall thickness is small, and thin plate approximations hold.
\end{itemize}
Since the moment acts across the entire cross-section of II, the stiffness of II can be calculated as $\tau_{II}=2EI/w$, where $E$, $I$, and $w/2$ are the Young's modulus, area moment of inertia and length of section II respectively. The moment on section I acts only at the corner (refer to point 2 in \cref{fig:Schem2}). To derive its stiffness, we consider this section in local coordinates ($xyz$) with its origin at 4, wherein the section lies horizontally along the $xy$ plane. The deformation of the plate due to the moment is approximated through suitable polynomials. The vertical displacement ($u_z$) along the $z$-axis is assumed to be of the form
\begin{equation}
    u_z=(\alpha+\beta y^2+\gamma y^4)(\mu x + \nu x^3),
\end{equation}
where $\alpha$, $\beta$, $\gamma$, $\mu$ and $\nu$ are constants. Here, only even terms of the polynomial are considered along $y$ to make the function even (due to mirror symmetry - refer to the red plane in \cref{fig:Schem1}) while it is assumed to be odd along $x$. The boundary condition on 3-4 implies that $(\partial u_z/\partial x)|_{x=0}=0$, implying that $\mu=0$. Therefore, with $a=\alpha\nu$, $b=\beta\nu$ and $a=\gamma\nu$, we have
\begin{equation} \label{eqn:w}
    u_z=(a+b y^2 + c y^4)(x^3)
\end{equation}
Further, since section II is assumed to be narrow, the deformation of section I must be such that 2-5 has 0 displacement along the X-axis. In other words, the deformation of 2-5 must occur along a plane parallel to the $YZ$ plane and containing 2-5. Note that in the local coordinate, this is a plane containing 2-5 and at an inclination of $\theta$ (about $x$) from the $yz$ plane. This implies that the displacement ($u_x$) along $x$ at $x=l$ must be
\begin{equation}
    u_x|_{x=l}=(u_z|_{x=l})\tan{\theta}
\end{equation}
We assume the variation of $u_x$ along $x$ to follow $u_z$ and therefore, we have
\begin{equation} \label{eqn:u}
    u_x=(u_z|_{x=l})\tan{\theta}(x^3/l^3)=u_z\tan{\theta}
\end{equation}
Now, assume that a moment $M$ at 2 (refer \cref{fig:Schem2}) causes an inclination of $\psi$ at 2 about the $X$-axis. Therefore, we have 
\begin{equation} \label{eqn:slopecondn}
    \frac{\partial u_z/\partial y}{\cos{\theta}}\bigg|_{(x=l,y=w/2)}=\psi.
\end{equation}
\Cref{eqn:slopecondn} can be used to eliminate a constant ($b$) from \cref{eqn:w} to yield
\begin{equation} \label{eqn:wf}
    u_z=(a+ \left(\dfrac{\psi\cos{\theta}}{w l^3}-\dfrac{c w^2}{2}\right) y^2 + c y^4)x^3
\end{equation}
Finally, the displacement ($u_y$) along $y$ is assumed to be linear along $y$ and follow a cubic variation similar to $u_z$ and $u_x$ along $x$ to give
\begin{equation} \label{eqn:v}
u_y=f y x^3,    
\end{equation}
where $f$ is a constant. The strain in the plate is given by
\begin{subequations} \label{eqn:epsilon}
\begin{equation}
    \epsilon_{xx}=\pdv{u_x}{x}-z\pdv[2]{u_z}{x}.
\end{equation}
\begin{equation}
    \epsilon_{yy}=\pdv{u_y}{y}-z\pdv[2]{u_z}{y}.
\end{equation}
\begin{equation}
    \epsilon_{xy}=\left(\pdv{u_x}{y}+\pdv{u_y}{x}\right)/2-z\pdv{u_z}{x}{y}.
\end{equation}   
\end{subequations}
Using \cref{eqn:u,eqn:v,eqn:wf} in \cref{eqn:epsilon}, we write the strains in the plate as follows
\begin{align}
    \epsilon_{xx}&=\frac{3 x (x \tan{\theta}-2 z) \left(l^3 w \left(2 a-c w^2 y^2+2 c y^4\right)+2 \psi  y^2 \cos{\theta}\right)}{2 l^3 w} \label{eqn:exx}\\
    \epsilon_{yy}&=x^3 \left(c z \left(w^2-12 y^2\right)+f-\frac{2 \psi  z \cos{\theta}}{l^3 w}\right) \label{eqn:eyy}\\
    \epsilon_{xy}&=\frac{x^2 y \left(l^3 w \left(-c x \tan{\theta} \left(w^2-4 y^2\right)+6 c z \left(w^2-4 y^2\right)+3 f\right)+2 x \psi  \sin{\theta}-12 \psi  z \cos{\theta}\right)}{2 l^3 w} \label{eqn:exy}
\end{align}
Neglecting the effect from the Poisson's ratio, the corresponding stresses are given by
\begin{subequations} \label{eqn:sigma}
\begin{equation}
     \sigma_{xx}=E \epsilon_{xx}.
\end{equation}
\begin{equation}
    \sigma_{yy}=E \epsilon_{yy}.
\end{equation}
\begin{equation}
    \sigma_{xy}=(E/2) \epsilon_{xy}.
\end{equation}   
\end{subequations}
Furthermore, the energy of an infinitesimal element of the section is given by
\begin{equation} \label{eqn:dut}
    dU=\frac{1}{2}\left(\sigma_{xx}\epsilon_{xx}+\sigma_{yy}\epsilon_{yy}+\sigma_{xy}\epsilon_{xy}\right)dxdydz.
\end{equation}
Therefore, using \cref{eqn:sigma} and \cref{eqn:exx,eqn:eyy,eqn:exy} in \cref{eqn:dut}, we obtain
\begin{dmath} \label{eqn:du}
    dU=\frac{E x^2}{16 l^6 w^2}\bigg(18 (x \tan{\theta}-2 z)^2 (l^3 w (2 a-c w^2 y^2+2 c y^4)+2 \psi  y^2 \cos{\theta} )^2
    +x^2 y^2 (l^3 w (c x \tan{\theta} (w^2-4 y^2)-3 (2 c z (w^2-4 y^2)+f))
    -2 x \psi  \sin{\theta}+12 \psi  z \cos{\theta})^2
    +8 x^4 (l^3 w (c z (w^2-12 y^2)+f)-2 \psi  z \cos{\theta})^2\bigg)dxdydz.
\end{dmath} 
The total energy ($U$) is found by integrating \cref{eqn:du} through the volume as $U=\int^{t_w/2}_{-t_w/2}{\int^{w/2}_0{\int^l_0{dU}}}$ to give
\begin{dmath} \label{eqn:U}
    U=\frac{E t_w}{11289600 l^3 w} \bigg(3 l^2 w^2 \big(\tan ^2{\theta} \big(1693440 a^2 l^6-98784 a c l^6 w^4+160 c^2 l^8 w^6+2247 c^2 l^6 w^8+10584 w^2 \psi ^2 \cos (2 \theta )+10584 w^2 \psi ^2\big)-56 l^3 w \tan{\theta} (70 c f l^4 w^3-9 \psi  \sin{\theta} (560 a-27 c w^4))+160 l^8 (840 f^2-c^2 w^6)+20 l^2 \sec ^2{\theta} \big(8 c^2 l^6 w^6+28 c l^3 w^3 \psi  \cos (3 \theta )+441 f^2 l^4 w^2 \cos (2 \theta )+441 f^2 l^4 w^2+245 f l^2 w \psi  \sin (3 \theta )-35 \psi ^2 \cos (4 \theta )+35 \psi ^2\big)-140 l^4 w \psi  \sec{\theta} (4 c l w^2-35 f \tan{\theta})\big)+7 t_w^2 \big(403200 a^2 l^6 w^2-23520 a c l^6 w^6+24 l^3 w^3 \psi  \cos{\theta} (2800 a-3 c w^2 (56 l^2+45 w^2))+3840 c^2 l^{10} w^6+576 c^2 l^8 w^8+535 c^2 l^6 w^{10}+9600 l^4 \psi ^2+5040 l^2 w^2 \psi ^2+120 \psi ^2 \cos (2 \theta ) (80 l^4+42 l^2 w^2+21 w^4)+2520 w^4 \psi ^2\big)\bigg)
\end{dmath}
Finally, the constants $a$, $c$, $f$, and $\psi$ are obtained through energy minimization with respect to these constants. In other words, the following set of equations is solved simultaneously
\begin{subequations} \label{eqn:MinEqns}
\begin{equation}
    \pdv{U}{a}=0
\end{equation} 
\begin{equation}
    \pdv{U}{c}=0
\end{equation} 
\begin{equation}
    \pdv{U}{f}=0
\end{equation} 
\begin{equation}
    \pdv{U}{\psi}=M
\end{equation} 
\end{subequations}
Using \cref{eqn:U} in \cref{eqn:MinEqns} and solving simultaneously using the symbolic computing toolbox in MATHEMATICA\textsuperscript{\textregistered} yields
\begin{dgroup}
    \begin{dmath}
        a= 525 M w^2 \sec{\theta} \Big(3 l^2 w^2 \tan ^2{\theta} (28800 l^4+48965 l^2 w^2+7056 w^4)
        -28 (160 l^2+21 w^2) (2400 l^4-81 l^2 w^2-20 w^4) t_w^2\Big)/G,
    \end{dmath}
    \begin{dmath}
        c=882000 M \sec{\theta} \Big(4 (63 l^2+20 w^2) (160 l^2+21 w^2) t_w^2
        +l^2 \tan ^2{\theta} (9600 l^4+23075 l^2 w^2+3024 w^4)\Big)/G,
    \end{dmath}
    \begin{dmath}
        f=1176000 M w^2 \tan{\theta} \sec{\theta} \Big(-7 (600 l^5+27 l^3 w^2+10 l w^4) t_w^2\\
        -9 l^3 w^2 \tan ^2{\theta} (5 l^2+14 w^2)\Big)/G,
    \end{dmath}
    \begin{dmath} \label{eqn:psi}
        \psi=25200 l^3 M \sec ^2{\theta}\Big(840 w (160 l^2+21 w^2) (20 l^4+3 l^2 w^2+w^4) t_w^2+l^2 w^3 \tan ^2{\theta} (96000 l^4+245945 l^2 w^2+31752 w^4)\Big)/G,
    \end{dmath}    
\end{dgroup}
where $G$ is given as
\begin{dmath*}
    G=E t_w \Big(9 l^4 w^4 \tan ^4{\theta} \left(240000 l^6+672875 l^4 w^2+215180 l^2 w^4+21168 w^6\right)+5 l^2 w^2 \tan ^2{\theta} \left(51840000 l^8+101026800 l^6 w^2+18697875 l^4 w^4+1043252 l^2 w^6+42336 w^8\right) t_w^2+140 \left(160 l^2+21 w^2\right) \left(72000 l^8+48600 l^6 w^2+13701 l^4 w^4+630 l^2 w^6+20 w^8\right) t_w^4\Big).
\end{dmath*}
It is evident from \cref{eqn:psi} that $\psi$ is linear with respect to M (Note that small strain assumptions have been made). Consequently, the stiffness of section I (and III equivalently) can thus be found as $\tau_{I}=M/\psi$, which yields
\begin{dmath} \label{eqn:tau1}
        \tau_I=G/\Big(25200 l^3 \sec ^2{\theta}\big(840 w (160 l^2+21 w^2) (20 l^4+3 l^2 w^2+w^4) t_w^2+l^2 w^3 \tan ^2{\theta} (96000 l^4+245945 l^2 w^2+31752 w^4)\big)\Big).
    \end{dmath}
Therefore, using \cref{eqn:tau1}, the total stiffness of the equivalent torsion spring can thus be written as
\begin{dmath} \label{eqn:tau}
        \tau=\tau_{II}+2G/\Big(25200 l^3 \sec ^2{\theta}\big(840 w (160 l^2+21 w^2) (20 l^4+3 l^2 w^2+w^4) t_w^2+l^2 w^3 \tan ^2{\theta} (96000 l^4+245945 l^2 w^2+31752 w^4)\big)\Big),
    \end{dmath}
where $\tau_{II}$ is given by
\begin{equation}
    \tau_{II}=\frac{E \left(d^3 \tan ^2{\theta} \sec{\theta} t_w+4 d \cos{\theta} t_w^3\right)}{24 w},
\end{equation}
where $d$ is the width of the bridge film. The above expression is found by evaluating the area moment of inertia of II as twice that of a rectangular section rotated through $\theta$. 

The stiffness of the equivalent torsion spring can thus be approximated by \cref{eqn:tau}. It is to be noted that the above expression has been derived assuming a certain polynomial form for the displacements. Consequently, it is found that the \cref{eqn:tau} holds only within a certain domain. Accordingly, the above expression is found to be valid for $(10\leq l/t_w\leq 60)$. A large class of hexagonal cellular structures fall within this domain, and the above formula can be used to approximate their stiffness. On the other hand, higher accuracies and larger domains of validity require higher-order approximations.
\bibliography{references}